\documentclass[12pt, preprint]{aastex}

\usepackage{amsmath}
\usepackage{amssymb}

\newcommand{\be}{\begin{equation}}
\newcommand{\ee}{\end{equation}}
\newcommand{\bea}{\begin{eqnarray}}
\newcommand{\eea}{\end{eqnarray}}

\shorttitle{Milky Way Models}
\shortauthors{Widrow & Pym}

\begin{document}

\title{Dynamical Models for Disk Galaxies with Triaxial Halos}

\author{Lawrence M. Widrow\altaffilmark{*}} \affil{Department of
Physics, Engineering Physics, and Astronomy, Queen's University,
Kingston, ON, K7L 3N6, Canada\\ \medskip (submitted to the
Astrophysical Journal, October, 2007)}
\altaffiltext{1}{widrow@astro.queensu.ca}

\begin{abstract}

  We construct self-consistent dynamical models for disk galaxies with
  triaxial, cuspy halos.  We begin with an equilibrium, axisymmetric,
  disk-bulge-halo system and apply an artificial acceleration to the
  halo particles.  By design, this acceleration conserves energy and
  thereby preserving the system's differential energy distribution 
  even as its phase space distribution function is altered.  The
  halo becomes triaxial but its spherically-averaged density profile
  remains largely unchanged.  The final system is in equilibrium, to a
  very good approximation, so long as the halo's shape changes
  adiabatically.  The disk and bulge are ``live'' while the halo is
  being deformed; they respond to the changing gravitational potential
  but also influence the deformation of the halo.  We test the
  hypothesis that halo triaxiality can explain the rotation curves of
  low surface brightness galaxies by modelling the galaxy F568-3.

\end{abstract}



\section{INTRODUCTION}

Dark matter halos -- at least the ones found in cosmological
simulations -- have a number of universal traits.  Most famously,
their density profiles have a shape that is nearly independent of
mass, formation epoch, and cosmological model \citep{nfw96}.  Their
angular momentum distribution \citep{bullock01}, phase space density
\citep{taylor01}, and velocity anisotropy \citep{hansen06} profiles
also appear to follow universal forms.  In other respects, halos are
rather diverse.  Simulated halos are typically triaxial with axis
ratios that range from $0.6$ to $1$ (See \citet{dubinski91, warren92}
and more recently, \citet{novak06}).  The shapes of real halos are
more difficult to determine but promising observational approaches do
exist.  Probes of the Galactic halo include flaring of the gas disk
\citep{olling} and tidal streams of satellite galaxies
\citep{johnston99}.  The shapes of halos in other galaxies can be
determined, at least statistically, by weak gravitational lensing
surveys \citep{hoekstra, mandelbaum, parker}.  On the other hand,
triaxiality can bias attempts to determine a halo's density profile
from the rotation curve of the disk that sits within it
\citep{hayashi06}.

Our main goal in writing this paper is to introduce a novel scheme to
generate self-consistent dynamical models for disk galaxies with
triaxial halos.  Our models can be tailored to fit observational data
for specific galaxies and therefore provide a testing ground to study
the disk-halo connection.  We consider the effects of halo triaxiality
on the rotation curves of low surface brightness galaxies (LSBs)
and briefly discuss other applications of the models.

Halos in simulations of a cold dark matter (CDM) universe have central
cusps with $\rho\propto r^{-\gamma}$ where $\gamma\simeq 1$
\citep{nfw96}.  In dark matter-dominated galaxies, this density
profile would seem to imply a rotation curve where $v\propto r^{1/2}$
as $r\to 0$.  By contrast, a halo with a constant density core implies
$v\propto r$ as $r\to 0$.  LSBs, which are believed to be dark-matter
dominated at small radii, have rotation curves that generally favor a
constant density core over a $\gamma=-1$ cusp.  This result represents
one of the most serious challenges to the CDM scenario \citep{moore94,
  flores94, mcg98} and has inspired some rather exotic alternatives.
De Blok \& McGaugh (1998), for example, suggested that LSB rotation
curves could be explained by Modified Newtonian Gravity while
\citet{firmani00} and \citet{mo00}) invoked dark matter
self-interactions to flatten the central cusp of the halo.

The connection described above between a galaxy's rotation curve and
the intrinsic density profile of its halo assumes that the halo is
spherically symmetric.  However, if a galaxy's halo is triaxial, then
gas in the disk will move on non-circular orbits and, under certain
conditions, the observed rotation curve will rise approximately
linearly even if the intrinsic halo density profile has a steep cusp.
\citet{hayashi06} and \citet{hayashi07} presented this argument as a
means of reconciling LSB rotation curves with the predictions of the
CDM model.

\citet{hayashi06} and \citet{hayashi07} derived model rotation curves
by calculating closed orbits in the potential generated by a triaxial
halo.  In this paper, we derive rotation curves by making
pseudo-observations of a disk that is embedded in the halo.
Deviations from axial symmetry in disk and halo are generated
concurrently and self-consistently.

A number of methods exist for constructing models of, and embedding
disks in, triaxial halos.  For example, \citet{moore04} show that the
remnant of a major merger between two equilibrium spherical halos is
triaxial.  \citet{bailin07} describe how to set up an equilibrium disk
in a combined halo-disk potential.  Our approach produces an N-body
galaxy complete with disk, bulge, and triaxial halo.  (Central black
holes may also be included, as in \citet{wid05}.)  It is inspired by
the method outlined in \citet{holley01}.  In that scheme, dubbed
``adiabatic squeezing'', particles of an equilibrium halo are
subjected to an artificial drag by modifying the force law of a
standard N-body code and evolving the system forward in time.  A
triaxial halo is created if the drag has a different strength along
three orthogonal directions, that is, along what become the three
principle axes of the halo.  The final model will be in equilibrium,
to a good approximation, so long as the timescale for the halo's shape
to change is slow as compared to the typical orbital timescale of the
system.

Adiabatic squeezing causes a halo to shrink in size.  For an isolated
halo, this shrinking can be reversed by simply rescaling the positions
and velocities of the particles.  Obviously, this method is unsuitable
for disk-bulge-halo systems since the disk and bulge would be
disrupted in an unphysical way.  We propose a modification of this
method in which drag is applied along one axis and ``negative drag''
is applied along the other two or vice versa depending on whether one
wants a prolate or oblate halo.  We require that for each particle,
the change in energy due to the artificial drag force is zero.  In
this way, we change the phase space distribution of the particles but
not their energy distribution.  As noted in \citet{BT}, {\it if two
  systems have the same energy distribution, their
  spherically-averaged density profiles will be very similar even if
  their phase space distribution functions are different.}

Our starting point is the equilibrium model of \citet{wid07} which
comprises a Sersic bulge, cuspy dark halo, and exponential disk.  The
model is described in terms of a phase space distribution function
(DF) which, in turn, is a function of the integrals of motion.  In the
current version of the model, the halo component of the DF depends
only on the energy.  In the absence of a disk, the halo is spherically
symmetric.  With the disk included, the halo is flattened slightly but
is still axisymmetric.  Adiabatic deformation allows us to extend our
disk-bulge-halo model to systems with triaxial halos.

In Section 2, we describe our method and construct an example of an
isolated triaxial halo.  We then consider the LSB galaxy F568-3.  In
Section 3, we present axisymmetric, equilibrium models for this galaxy
based on its published surface brightness profile and circular speed
curve.  In Section 4, we show how transforming the axisymmetric halo
in one of these models into a triaxial halo changes the shape of the
rotation curve.  We conclude in Section 5, by summarizing our results
and briefly discussing further applications of the method.

\section{METHOD}

We begin with an N-body equilibrium halo and evolve the system forward
in time using standard techniques augmented by an artificial,
energy-conserving acceleration.  To be precise, we introduce an
acceleration into the equations of motion given by
\begin{equation}
a_x = \frac{\left (\beta_1-\beta_2\right )v_y^2 + \beta_1 v_z^2}{v^2}\,v_x~,
\label{eq:ax}
\end{equation}
\begin{equation}
a_y = \frac{\left (\beta_2-\beta_1\right )v_x^2 + \beta_2 v_z^2}{v^2}\,v_y~,
\label{eq:ay}
\end{equation}
and
\begin{equation}
a_z = -\frac{\beta_1v_x^2 + \beta_2 v_y^2}{v^2}\,v_z
\label{eq:az}
\end{equation}
where $v$ is the speed of the particle and ${\bf a}\cdot {\bf v} = 0$,
as required.  The coefficients $\beta_1$ and $\beta_2$ are
time-dependent.  Following \citet{holley01} we assume that $\beta_{1}$
and $\beta_2$ ``turn on'' at $t=0$ and increase to their respective
maximum values over a period $T_G$ with a time-dependence given by
\begin{equation}\label{eq:betatimedependence}
  \beta_i = 
\beta_{i,max}\left (3\left (t/T_G\right )^2
    - 2\left (t/T_G\right )^3\right )~.
\end{equation}
$\beta_1$ and $\beta_2$ remain constant for a time $T_C$ before
decreasing to zero over a time $T_D$.

A few comments regarding the parameters $\beta_1$, $\beta_2$, $T_G$,
$T_C$, and $T_D$ are in order.  First, the degree by which the halo
departs from spherical symmetry is given, roughly, by the integral
$\int \beta_i dt = \beta_i\left (T_G + T_C + T_D\right )$.  Rescaling
the $\beta$'s by a factor $f$ and the $T$'s by a factor $f^{-1}$
leaves the final halo shape unchanged with one important caveat.
$\beta^{-1}$ sets the timescale over which the halo's shape changes
and therefore must be longer than its dynamical time $\sim
a_h/\sigma_h$ in order to maintain adiabaticity.

In the case of an isolate halo equations \ref{eq:ax}-\ref{eq:az} admit
several discrete symmetries.  For example, interchanging $\beta_1$ and
$\beta_2$ is equivalent to interchanging $x$ and $y$.  Similar
symmetries are listed in Table 1.  The presence of a disk breaks these
symmetries.  In general, increasing $\beta_1$ causes the system to expand
along the $x$-axis while increasing $\beta_2$ causes the system to expand
along the $y$-axis.  Models with $\beta_1=\beta_2$ are axisymmetric about
the $z$-axis.

As an illustration, we transform an isolated, spherically-symmetric
halo into one that is triaxial.  We begin with a halo 
whose density profile is given by
\begin{equation}\label{eq:haloprofile}
\tilde{\rho}_{\rm halo} = \frac{2^{2-\gamma}\sigma_h^2}{4\pi a_h^2}
\frac{1}{\left (r/a_h\right )^{\gamma}
\left (1 + r/a_h\right )^{3-\gamma}}\,
{\rm erfc}\left (\frac{r-r_h}{\sqrt{2}\delta r_h}\right )~.
\end{equation}
For this example, we set $\gamma = 1$ (the NFW value), $a_h= 10\,{\rm
  kpc}$, $\sigma_h=100\,{\rm km\,s^{-1}}$, $r_h = 100\,{\rm kpc}$ and
$\delta r_h = 10\,{\rm kpc}$.  We choose $\beta_{1,max} = 0.24 {\rm \,
  Gyr}^{-1}$, $\beta_{2,max} = 0.08 \,{\rm Gyr^{-1}}$, $T_G = T_D =
1\,{\rm Gyr}$ and $t_c=3\,{\rm Gyr}$.  As required, the characteristic
timescale for the halo, $a_h/\sigma_h = 100\,{\rm Myr}$, is short
compared to the timescale, $\beta^{-1} = 4-12\,{\rm Gyr}$, associated
with the artificial force.  Our choice for the $\beta_i$ yields a halo
whose short axis is along the $z$-direction and whose long axis is
along the $x$-direction.

The model is evolved forward in time using the N-body code from
\citet{stiff03} which is based on the algorithm described in
\citet{dehnen00}.  The code uses a multipole expansion for cell-cell
interactions; computational costs scale approximately linearly with
particle number $N$.  The softening length is $200\,{\rm pc}$ and the
timestep is $1\,{\rm Myr}$.  The system is evolved for a period of
$15\,{\rm Gyr}$.

\begin{figure}
\epsscale{1.0}
\plotone{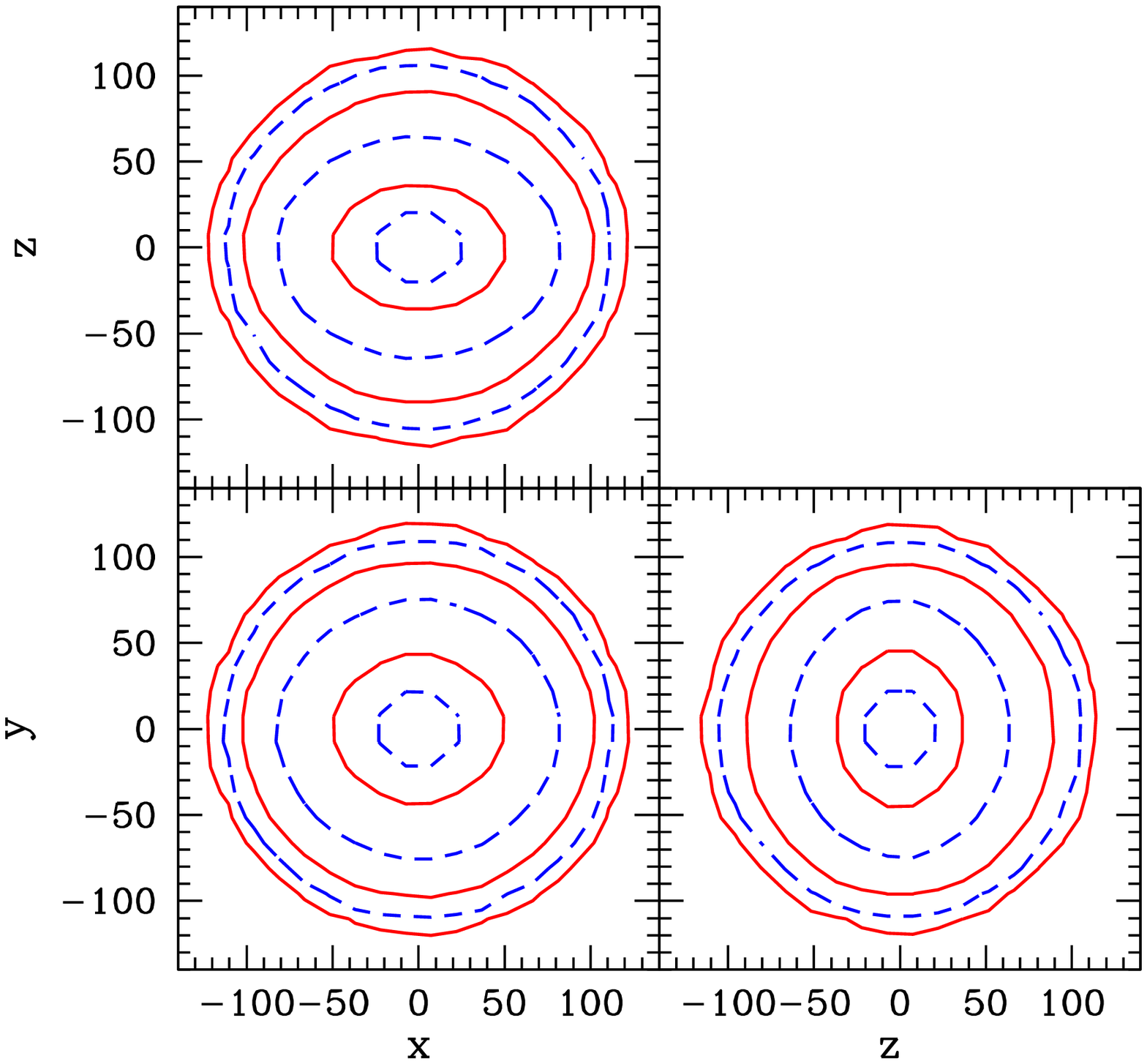}
\caption{Surface density contours along the three principle axes for the
triaxial isolated halo constructed in Section 2.
Spacing between solid contours is 1 dex.}
\label{fig:xyz}
\end{figure}

\begin{figure}
\epsscale{1.0}
\plotone{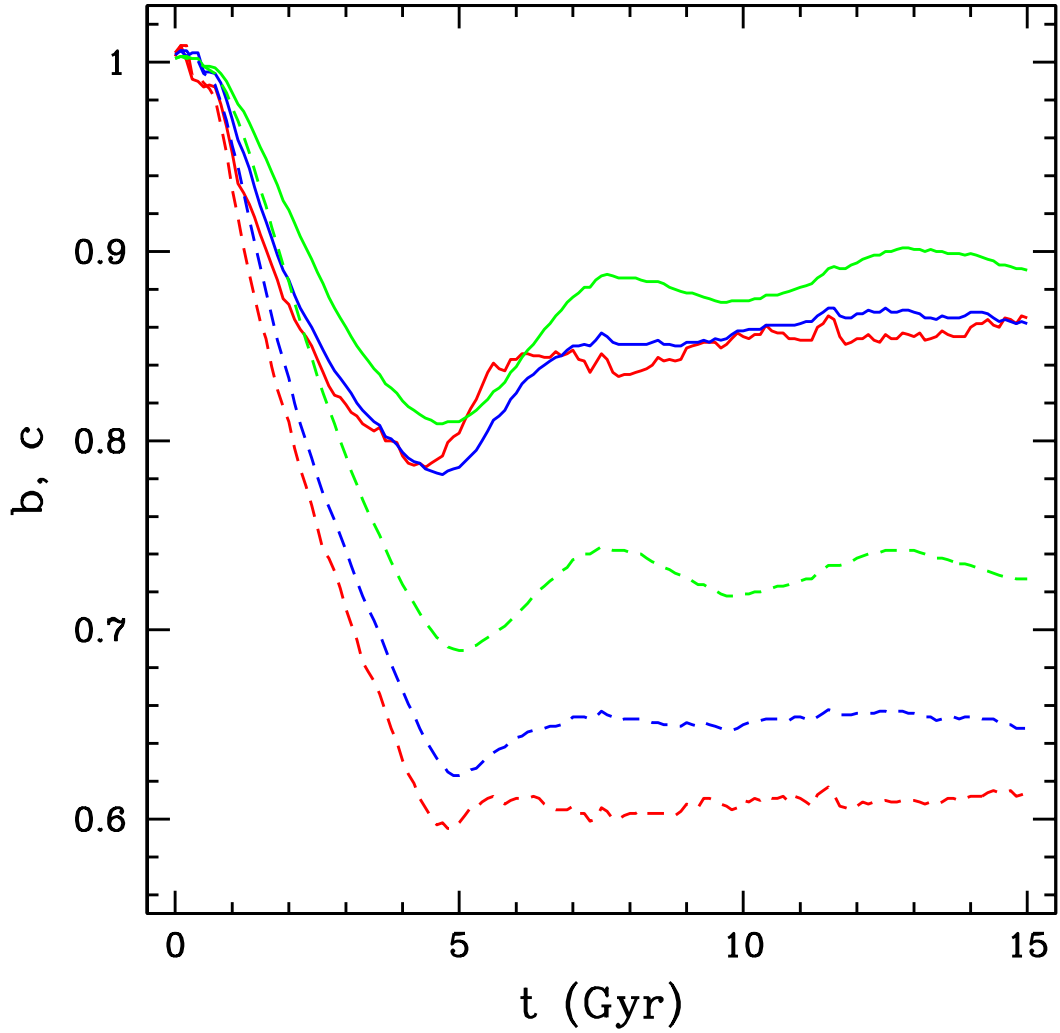}
\caption{Axes ratios as a function of time.  Solid curves show $b$;
  dashed curves show $c$.  Red, blue, and green curves are for,
  respectively, the inner one third of the particles, the inner
two thirds of the particles, and the entire halo.}
\label{fig:axes}
\end{figure}

In Figure \ref{fig:xyz}, we show a contour plot of the projected
surface density of the halo along the three principle axes.  Note that
the departure from spherical symmetry is strongest in the inner parts
of the halo.  In Figure \ref{fig:axes}, we show the axis ratios as a
function of time.  To be precise, we model the density field
as an ellipsoidal distribution,
\begin{equation}
\rho = \rho(\tilde{r})~~~~{\rm where}~~~~
\tilde{r} = x^2 + \frac{y^2}{b^2} + \frac{z^2}{c^2}~.
\end{equation}
The parameters $b$ and $c$ are calculated through an iterative
procedure as outlined in \citet{dubinski91}.  We show the results
using the inner third of the particles, the inner two thirds of the
particles, and all of the particles.  Again, we see that the halo is
more spherical in the outer parts.  Note that the axis ratios at all
radii oscillate a bit at $t=5\,{\rm Gyr}$, the time when the
artificial acceleration is turned off.  After this time, the axis
ratios in the inner two thirds of the halo quickly settle down to
constant values.  The oscillations damp more slowly in the outer 
parts of the halo where the dynamical time is not much shorter than
$\beta_i^{-1}$.  One can minimize the oscillations by increasing
the $T's$ and decreasing the $\beta$'s but at the cost of additional
computation time.

\begin{figure}
\epsscale{1.0}
\plotone{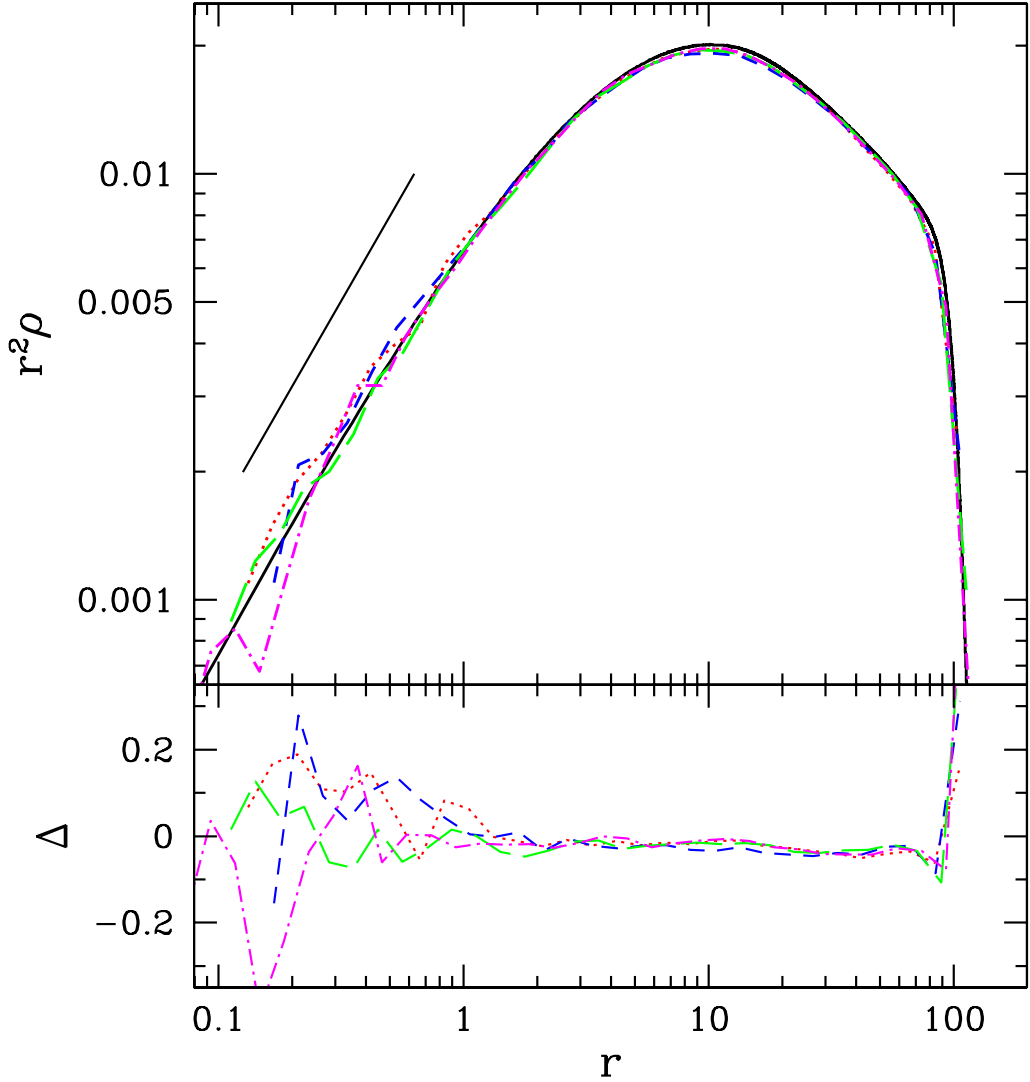}
\caption{Differential mass profile, $dM/dr\propto r^2\rho$ as a
  function of spherical radius $r$.  Line types are: analytic profile
  -- solid black curve; initial profile derived from the N-body
  distribution -- dot-dashed magenta curve; profile at $6\,{\rm Gyr}$
  -- dotted red curve; profile at $12\,{\rm Gyr}$ -- dashed blue
  curve.  The profile for the control experiment (no artificial
acceleration) at $12\,{\rm Gyr}$ is
  shown by the long-dashed green curve.  The straight solid black line
  corresponds to $r^2\rho\propto r$ or $\rho(r)\propto
  r^{-1}$.  Lower panel gives the fractional difference between the 4
  measured profiles and the the analytic expression (i.e., $\left
    (\rho_{\rm measured}-\rho_{\rm exact}\right )/\rho_{\rm exact}$.}
\label{fig:density}
\end{figure}

In Figure \ref{fig:density}, we show the spherically-averaged
differential mass profile, $dM/dr \propto r^2\rho$ for the initial
model and for the deformed model at $t=6\,{\rm Gyr}$ and $t=12\,{\rm
  Gyr}$ and compare with equation 5.  Also shown is the density
profile calculated from the initial conditions and the density profile
for the system evolved to $12\,{\rm Gyr}$ with no artificial
acceleration.  The former illustrates the role mass resolution plays
on the measured density profile while the latter illustrates the
effects of force softening and two-body relaxation.  We see that the
spherically-averaged density profile is preserved to within the
fluctuations introduced by these other effects.

Table 2 presents results for the axes ratios for other choices of
$\beta_1$ and $\beta_2$.  Note that models 1a-c (and likewise models
2a-b and models 3a-c) are equivalent through the symmetries described
in Table 1.

\section{THE LSB GALAXY F568-3}

In this section, we construct axisymmetric, equilibrium models for
F568-3, an LSB galaxy which has appeared in a number of studies.  We
describe our general axisymmetric disk-bulge-halo models, review
published photometric and kinematic observations for this galaxy, and
discuss the statistical techniques used to tailor the model to the
data.

\subsection{\it Equilibrium Models for Disk-Bulge-Halo Systems}

Our starting point is the dynamical galactic model described in
\citet{wid07}.  The model is axisymmetric and comprises an exponential
disk, a Sersic bulge, and a halo whose density profile is given by
equation \ref{eq:haloprofile}.  DFs for the bulge and halo are
functions of the energy, $E$, and constructed via an Abel integral
transform.  The DF for the disk is constructed from three integrals of
motion following the method outlined in \citet{kui95}.  The total DF
for the composite system self-consistently satisfies the collisionless
Boltzmann and Poisson equations.

The bulge has a spherically-averaged density profile given, to 
a good approximation, by
\begin{equation}\label{eq:prugnielsimien}
\tilde{\rho}_{\rm bulge}(r) = \rho_b\left (\frac{r}{R_e}\right )^{-p}
e^{-b\left (r/R_e\right )^{1/n}}~.
\end{equation}
This density profile yields the Sersic law,
\begin{equation}\label{eq:sersic}
\Sigma(r) = \Sigma_0 e^{-b\left (R/R_e\right )^{1/n}}~,
\end{equation}
for the projected mass density provided one sets $p = 1 - 0.6097/n +
0.05563/n^2$ \citep{ps97,ter05}.  $\Sigma_0$, $R_e$ and $n$ are free
parameters while the constant $b$ is adjusted so that $R_e$ encloses
half the total projected light or mass.

The disk DF depends on $E$, the angular momentum about the symmetry
axis, $L_z$, and an approximate integral of motion, $E_z$, which
corresponds to the energy associated with vertical motions of stars in
the disk.  The DF is adjusted so that the intrinsic three-dimensional
density distribution and velocity dispersion profile are given,
respectively, by
\begin{equation}
\rho_{\rm disk}\left (R,\,z\right ) = \rho_0 \exp^{-R/R_d}\,
{\rm sech}^2{\left (z/z_d\right )}
{\rm erfc}\left (\left (R-R_{\rm out}\right )/\delta R_{\rm out}\right )~.
\end{equation}
 and
\begin{equation}\label{eq:radialdispersion}
\sigma_R^2(R) = \sigma_{R0}^2 \exp{\left (-R/R_\sigma\right )}~.
\end{equation}

\subsection{\it Surface Brightness Profile and Rotation Curve for  F568-3}

Multi-band photometry for the LSB galaxy F568-3 is presented in
\citet{deblok95}.  The galaxy resembles a normal late-type galaxy
exhibiting a disk and faint spiral arms.  However, its central B-band
surface brightness is more than a magnitude fainter than the Freeman
value \citep{freeman70} placing it squarely in the category of LSBs.
For the purpose of modelling the galaxy, we use the R-band surface
brightness profile from Figure 2 of \citep{deblok95}.

High-resolution rotation curves for F568-3 are described in
\citet{mcgaugh01}.  The circular speed rises approximately linearly to
$80\,{\rm km\,s^{-1}}$ within $4\,{\rm kpc}$.  It continues to rise
beyond this radius reaching a maximum value of $\sim 100\,{\rm
  km\,s^{-1}}$ at $R\simeq 12\,{\rm kpc}$.

\subsection{\it Markov Chain Monte Carlo Analysis of F568-3}

We use Bayesian statistics and the Markov Chain Monte Carlo (MCMC)
method to find suitable axisymmetric models for F568-3.  MCMC
provides an efficient means of mapping out the likelihood function
over the full multi-dimensional parameter space and has a number of
advantages over traditional maximum likelihood techniques.

For a particular choice of model parameters, one can construct a
likelihood function which quantifies the agreement between the model
and the data.  Maximization techniques, such as the simplex algorithm,
allow one to hone in on the ``best-fit'' model.  However, with a large
number of parameters, the likelihood function may become difficult to
characterize with many false maxima.  Moreover, the computational
costs of simple algorithms, such as grid-based searches, become
prohibitive.

The goal of our MCMC analysis is to calculate the posterior
probability density function, $p(M|D,I)$, of a Galactic model, $M$,
given data, $D$, and prior information, $I$.  From Bayes' theorem we
have
\begin{equation}\label{eq:bayes}
p(M|D,I) = \frac{p(M|I)p(D|M,I)}{p(D|I)}
\end{equation}
where $p(M|I)$ is the prior probability density and $p(D|I)\equiv \int
d{\bf A}\, p(M|D,I)$ is a normalization factor.  In MCMC, one
constructs a sequence or chain of models through parameter space
chosen according to a prescribed algorithm.  The distribution of
models along the chain will be proportional to $p\left (M|D,I\right )$
provided the chain is sufficiently long.

In this work, we use the Metropolis-Hastings algorithm \citep{met53,
  hast70} as outlined in \citet{gre05}.  The first model in the chain
is chosen at random.  A candidate for the second model is chosen by
taking a step in parameter space according to a proposal distribution.
Let ${\cal R}$ be the ratio of the likelihood function of the
candidate to that of the first model.  The candidate is accepted a
fraction, $f$, of the time where $f=\min\{1,{\cal R}\}$.  Otherwise,
the second model is taken to be identical to the first model.  The
process is repeated to find the third model and so forth.

Care must be taken in selecting a proposal distribution.  If the step
size is too short, the chain moves slowly through parameter space and
the time required to fully explore parameter space becomes
prohibitively large.  On the other hand, if the typical step size is
too large, the acceptance rate will be very low.  We use an iterative
approach, as outlined in \citet{wid07}, to choose an efficient
proposal distribution.

Since the data used in this work do not include observations of the
velocity dispersion, the parameters $\sigma_{R0}$ and $R_\sigma$ are
superfluous and may be ignored in fitting the galaxy.  Likewise, only
the major-axis surface brightness profile is used and therefore the
disk scale-height parameter, $z_d$, may be fixed to a reasonable
value.  Finally, $r_h$ may be set to any value greater than $15\,{\rm
  kpc}$ (i.e., beyond the outermost point of the observed rotation
curve) and $\delta r_h$ may be ignored.  The DF is thus specified by
ten free parameters.

Our set of model parameters must include the mass-to-light ratios of
the disk and bulge.  In general, the rotation curve fit for LSBs is
improved by choosing a very large mass-to-light ratio for the disk,
that is, by devising a galactic model that is disk-dominated in the
inner regions.  However, the required mass-to-light ratios are
typically unphysical.  Indeed, one can constrain mass-to-light ratios
using population synthesis models and galaxy colours \citep{bell01,
  bell03}.  In a Bayesian analysis such as MCMC, these constraints are
implemented through prior probabilities for the mass-to-light ratios.
For simplicity, we assume that these prior probabilities follow a
log-normal distribution.  Using the $B-R$ and $B-V$ profiles from
\citet{deblok95} and the color-M/L relations from \citet{bell03} we
find
\begin{equation}
{\rm log}{\left (\left (M/L\right )_{\rm disk}\right )} =  0.03 \pm 0.25
\end{equation}
and 
\begin{equation}
{\rm log}{\left (\left (M/L\right )_{\rm bulge}\right)} =  0.2 \pm 0.25
\end{equation}
The errors, which translate directly into the width of the prior
probability distribution, are meant to incorporate uncertainties in
the relations from \citet{bell03}, uncertainties in the colors, and
differences in the $M/L$-values obtained by using either $B-R$ or
$B-V$ colours.

\begin{figure}
\epsscale{.9}
\plotone{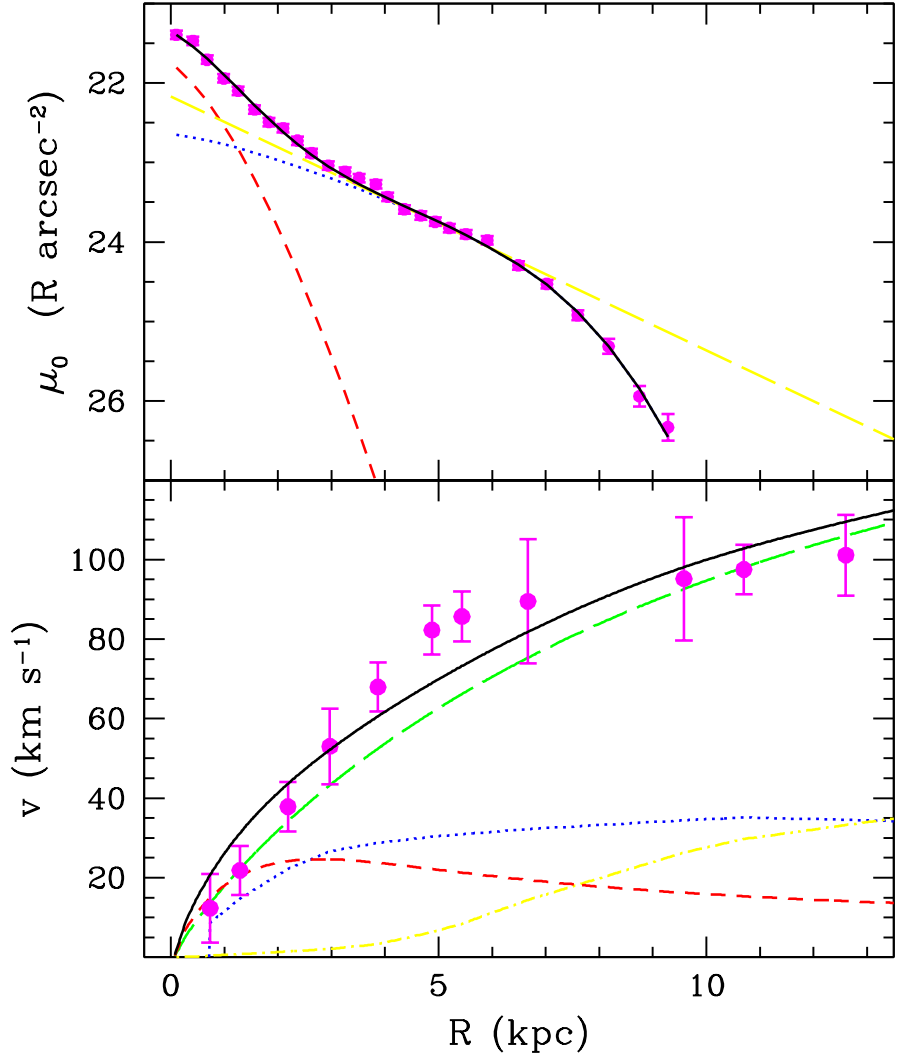}
\caption{Comparison of data with predictions for a typical model from
  the MCMC run where $\gamma$ is a free parameter.  Top panel shows
  surface brightness profile.  Observations from \citet{deblok95} are
  indicated by magneta dots.  Line types are as follows: total model
  surface brightness profile -- solid black curve; disk contribution
  -- blue dotted curve; bulge contribution -- red dashed curve;
  exponential disk model from \citet{deblok95} -- yellow long-dashed
  curve.  Bottom panel shows the rotation curve.  Observations from
  \citet{mcgaugh01} are indicated by magneta dots.  Total model --
  solid black curve; disk contribution -- blue dotted curve; bulge
  contribution -- red dashed curve; halo contribution -- green
  long-dashed curve; gas contribution -- yellow dot-dashed curve.}
\label{fig:profiles_free}
\end{figure}

\begin{figure}
\epsscale{1.0}
\plotone{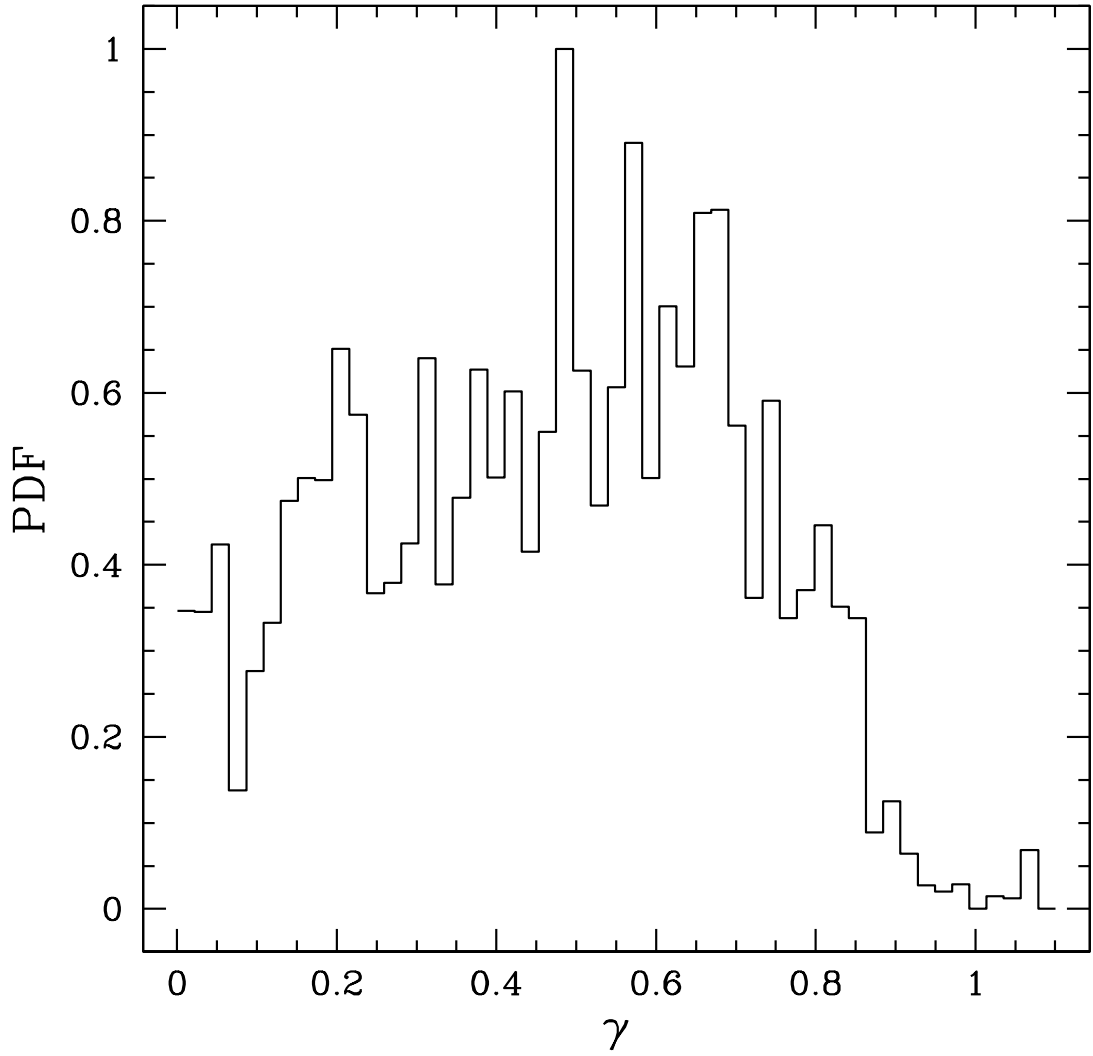}
\caption{Probability distribution function for $\gamma$ from 
the first MCMC where $\gamma$ is a free parameter.}
\label{fig:gamma}
\end{figure}

Two MCMC runs are conducted, one in which $\gamma$ is a free parameter
and one in which $\gamma$ is fixed to the NFW-value (i.e.,
$\gamma=1$).  The surface brightness profile and rotation curve fits
for a typical model from the first run are shown in Figure
\ref{fig:profiles_free}.  Also shown is the fit found by
\citet{mcgaugh01} which assumes an exponential disk and does not
include a bulge.  Evidently, an excellent fit to the full surface
brightness profile can be obtained provided both disk truncation and a
bulge are included in the model.  In Figure \ref{fig:gamma} we show
the probability distribution function for $\gamma$.  Clearly, the data
favor values of $\gamma$ between $0$ and $0.8$.

\begin{figure}
\epsscale{0.9}
\plotone{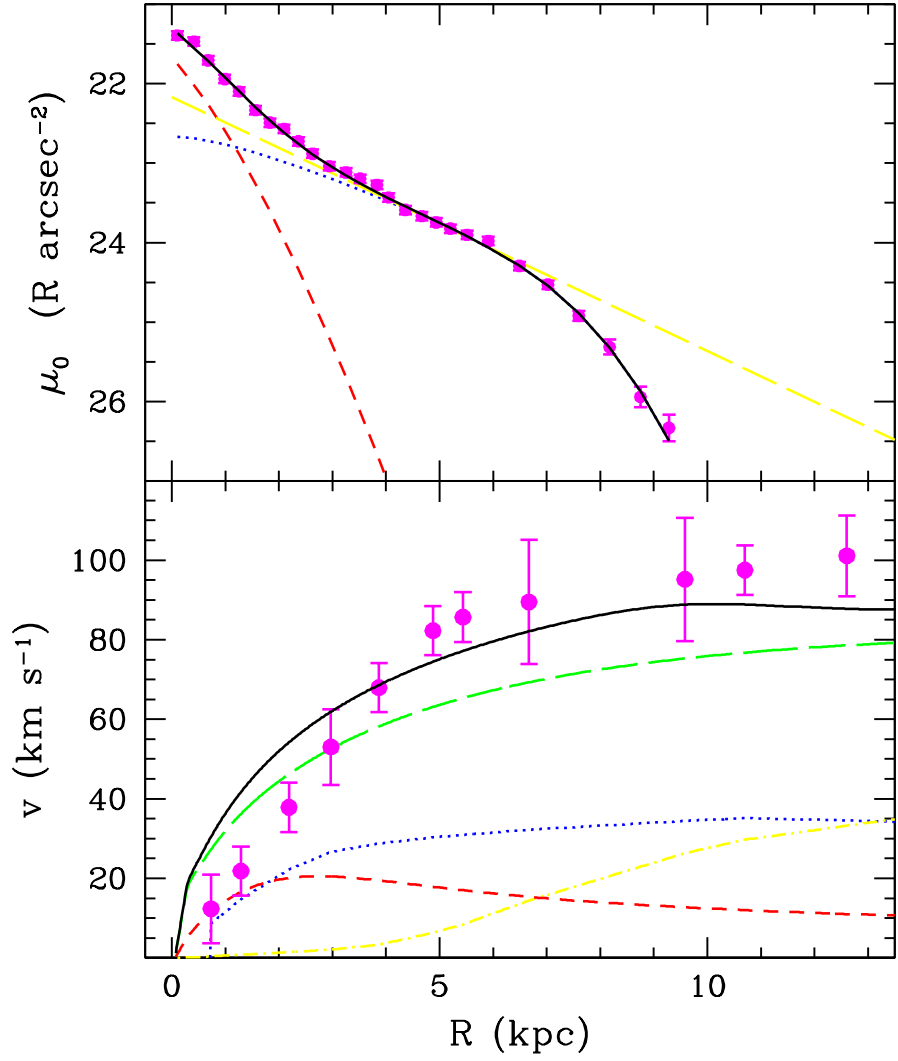}
\caption{Surface brightness profile and circular speed curve
for a typical model from MCMC chain where $\gamma=1$.
Line types are the same as in Figure \ref{fig:profiles_free}.}
\label{fig:profiles_fixed}
\end{figure}

The surface brightness profile and circular speed curve for a typical
model from our MCMC analysis with $\gamma=1$ is shown in Figure
\ref{fig:profiles_fixed}.  We find that the model rotation curve rises
too quickly as compared with the data, in agreement with previous
studies \citep{moore94, flores94, mcg98, deblok01}.

\section{MODELLING F-583 WITH A TRIAXIAL HALO}

Using the method outlined in Section 2, we transform the halo in one
of our axisymmetric $\gamma=1$ models.  We begin by generating an
N-body representation of the model with $500K$ particles for the halo,
$400K$ particles for the disk, and $100K$ particles for the bulge.  We
produce two examples of models with triaxial halos: Model I where
$\beta_1=0.24\,{\rm Gyr^{-1}}$ and $\beta_2=0.08\,{\rm Gyr^{-1}}$
(i.e., same choise of parameters as in our isolated halo example) and
Model II where $\beta_1 = 0.16\,{\rm Gyr^{-1}}$ and $\beta_2 =
-0.08\,{\rm Gyr^{-1}}$.  Note that the artificial acceleration is applied
only to the halo particles.

First, consider Model I.  Recall that in the example from Section 2
the short axis of the halo is aligned with the $z$-direction while the
long axis is aligned with the $x$-direction.  In a disk-bulge-halo
system, the same choice of parameters leads to a rather mild deviation
from axial symmetry since the intermediate axis is in the disk plane.

For an isolated halo, the choice of parameters used in Model II amounts
to a trivial interchange of the $y$ and $z$ coordinates.  In the
presence of the disk, the choice yields a halo model in which both the
long and short axes are in the disk plane and hence the departure from
axial symmetry is very strong.

\begin{figure}
\epsscale{1.0}
\plotone{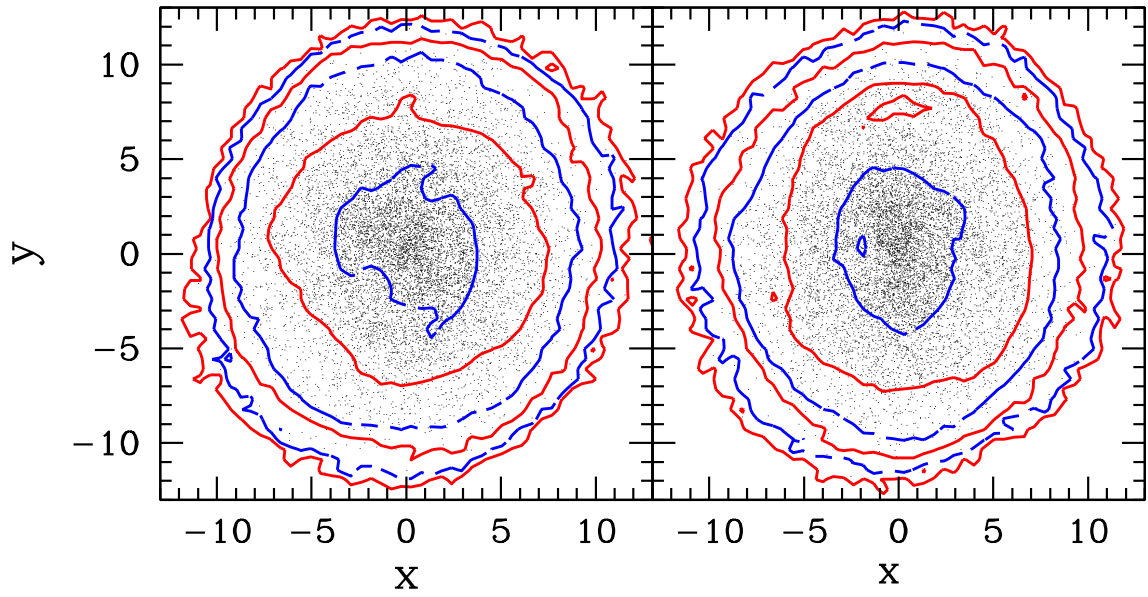}
\caption{Superposition of surface density contours and projected
particle distribution for Models A (top) and B (bottom).}
\label{fig:disk}
\end{figure}

In Figure \ref{fig:disk}, we show contour plots of the disk surface
density for Models I and II.  As expected, departures from circular
symmetry are more evident in Model II.  Also, as
expected, the long axis of the disk is perpendicular to the long axis
of the halo \citet{hayashi06}.

\begin{figure}
\epsscale{1.0}
\plotone{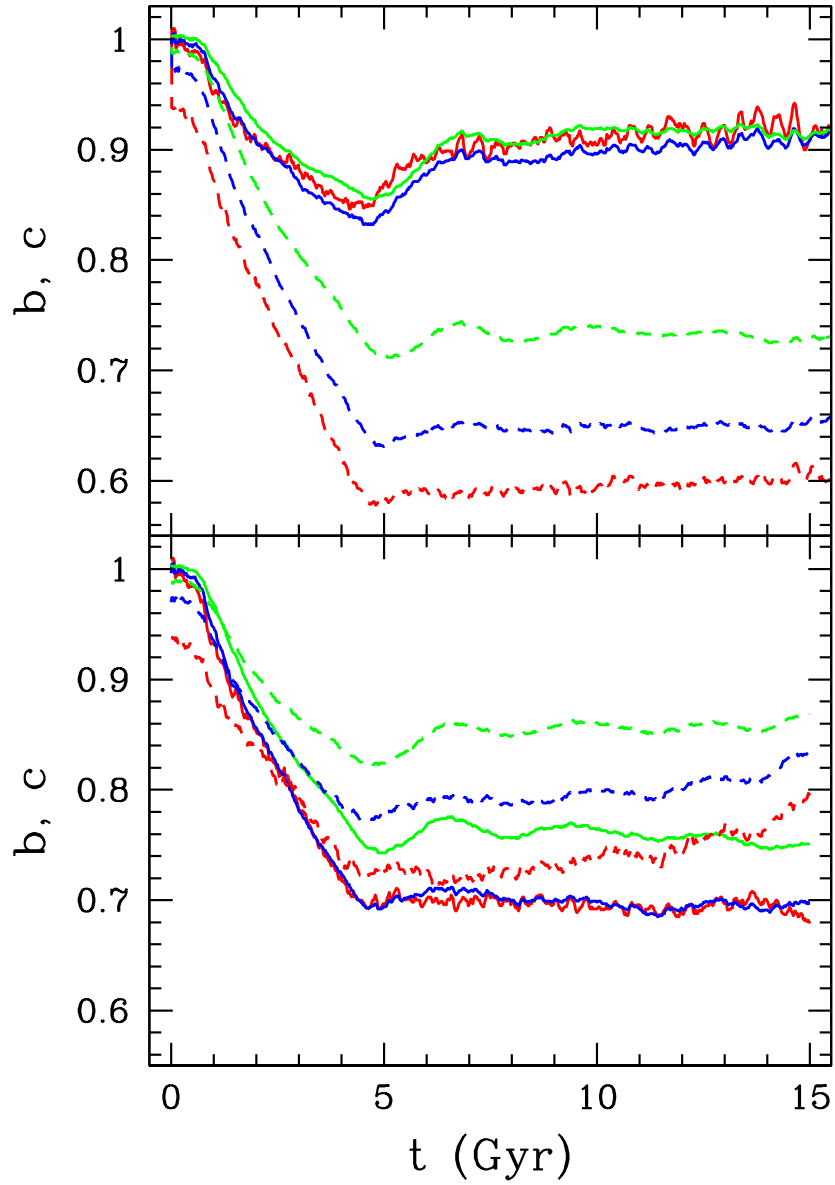}
\caption{Halo axis ratios as a function of time for Models A (top) and
  B (bottom).  Line types are the same as in Figures \ref{fig:axes}.}
\label{fig:LSB_axes}
\end{figure}

In Figure \ref{fig:LSB_axes} we show the evolution of the halo axes
ratios as a function of time for Models I and II.  The evolution of the
axis ratios in Model I is very similar to that found for the isolated
halo in Section 2.  By contrast, the influence of the disk is readily
evident in Model II;  the halo is flattened
along the $z$-direction and somewhat rounder in the $x-y$ plane than
it would be in the absence of the disk.  The end result is a halo that
is prolate with approximate axial symmetry about the $x$-axis.

\begin{figure}
\epsscale{1.0}
\plotone{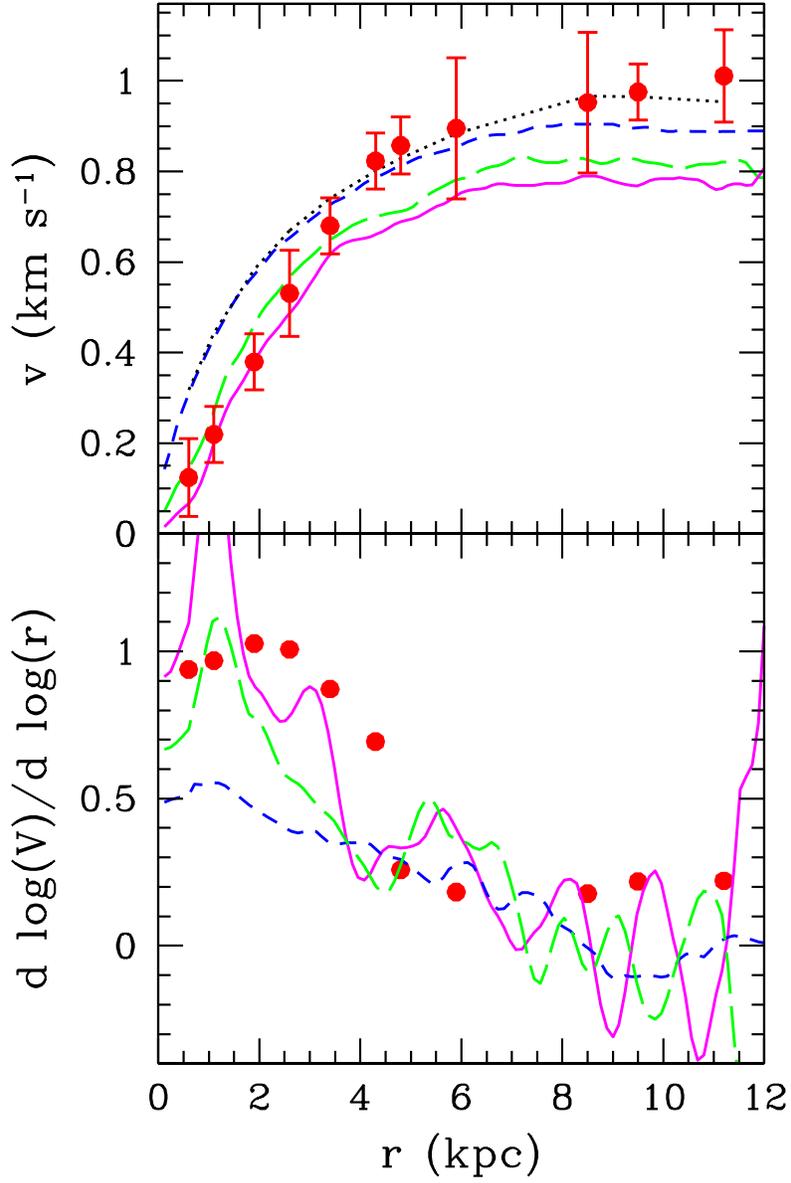}
\caption{Observed and model rotation curves.  Top panel shows the
  rotation curves; bottom panel shows the logarithmic slope.
  Observations from \citet{mcgaugh01} are shown as red dots.  Line
  types are as follows: black dotted line -- rotation curve for the
  initial model as derived from the potential; blue dashed line --
  rotation for the initial model as derived by measuring bulk motion of
  the disk stars; green long-dashed line -- rotation curve for Model
  I; magenta solid line -- rotation curve for Model II.}
\label{fig:LSB_vcirc}
\end{figure}

In Figure \ref{fig:LSB_vcirc} we show the rotation curves for Models I
and II as calculated along a slit placed on the major axis of the disk.
In both experiments, the maximum rotation speeds decreases relative to
their initial values by about $10\,{\rm km\,s^{-1}}$.  One might
imagine an iterative procedure in which, given these results, one
adjusts the initial model so that the final system better reproduces
the data.

Next, we consider the change in shape of the rotation curve produced
by the deformation of the halo.  In both experiments, the rotation curve
rises more slowly than in the initial, axisymmetric model.  Figure
\ref{fig:LSB_vcirc} shows the logarithmic slope of the rotation curve
and illustrates this point quantitatively.  We can also quantify the
change in shape of the rotation curve by considering the fitting
formula
\begin{equation}\label{courteaufit}
v(r) = v_0 \frac{1}{\left (1 + x^\alpha\right )^{1/\alpha}}
\end{equation}
where $x = a_v/r$ \citet{courteau97}.  $a_v$ and $v_0$ are scale
parameters while $\alpha$ dictates the shape of the function.
(\citet{courteau97} actually proposed a more elaborate fitting formula
but for our purposes, this form will suffice (see, for example,
\citep{hayashi04}.))  A cored-isothermal sphere yields a rotation
curve with $\alpha\ga 2$ while an NFW-halo yields a rotation curve
with $\alpha\simeq 0.6$ \citep{courteau97, hayashi04}.  We find the
following values for $\alpha$: observed rotation curve -- $5.35$;
initial, axisymmetric model -- $0.91$; Model I -- $1.94$; and Model II
-- $3.57$.  Clearly, Model II, where departures from axial symmetry
are strongest, comes closest to reproducing the shape of the rotation
curve.

\section{SUMMARY AND CONCLUSION}

The adiabatic squeezing method \citet{holley01} produces triaxial
halos that have shrunk in size and therefore requires that the
positions and velocities of the particles be rescaled.  This awkward
step precludes the technique from being applied to compound systems.
Our approach avoids this problem by using an {\it energy-conserving}
artificial force to deform the halos.

Our analysis of the LSB galaxy F568-3 begins with a discussion of
axisymmetric models.  We attempt to fit both photometric and kinematic
observations using Bayesian statistics and the MCMC method.  Our
excellent fit of the surface brightness profile requires a bulge and
disk truncation, neither of which were included in previous studies.
As for the rotation curve, we find that constant density cores do
better than density cusps in agreement with earlier studies of LSBs.

The second stage of our analysis is to deform the halo of a compound
system.  In agreement with \citet{hayashi06}, we show that the
rotation curve of F568-3 may indicate the presence of a triaxial halo
rather than a problem with the standard CDM model of structure
formation.  \citet{hayashi06} and \citet{hayashi07} construct rotation
curves by finding closed orbits in the gravitational potential of a
triaxial halo.  We calculate the rotation curves by making
pseudo-observations of a disk that is self-consistently embedded in a
dark halo.

There are two improvements that will add a further level of realism to
the analysis: the inclusion of a gas disk in the galactic models and
an iterative scheme whereby the initial model and artificial
acceleration parameters are adjusted so that the final model fits the
data in detail.  These improvements will be considered in a future
publication.

Our triaxial models have a wide range of applications.  For example,
they can be used to study the effect a non-spherical halo has on the
morphology of tidal streams from satellite galaxies and flaring and
warping of the gas disk.  The method can also be applied to bulges
where departures from axial symmetry are thought to be important.

\acknowledgements{It is a pleasure to thank J. Bailin, S. Courteau,
  J. Dubinski, S. McGaugh, and D. Puglielli for useful conversations.
  This work was supported by a grant from the Natural Sciences and
  Engineering Research Council of Canada.}


\newpage

\newpage

\newpage

\begin{deluxetable}{ccccccc}
\tablewidth{0pt}
\tablecaption{Invariances of Equations 1-3}
\tablehead{
\multicolumn{2}{c}{$\beta$} & \multicolumn{3}{c}{Coordinates} & \multicolumn{2}{c}{Axis ratios} \\}
\startdata
$\beta_1\to \beta_2$ & $\beta_2\to\beta_1$ & $x\to y$ & $y\to x$ & $z\to z$ 
& $b\to 1/b$ & $c\to c/b$\\
$\beta_1\to \beta_1-\beta_2$ & $\beta_2\to-\beta_2$ & $x\to x$ & $y\to z$ & $z\to x$ 
& $b\to c$ & $c\to b$\\
$\beta_1\to \beta_2-\beta_1$ & $\beta_2\to-\beta_1$ & $x\to y$ & $y\to z$ & $z\to x$ 
& $b\to 1/c$ & $c\to b/c$
\enddata
\end{deluxetable}

\begin{deluxetable}{ccccccc}
\tablewidth{0pt}
\tablecaption{Axis Ratios for Various Choices of $\beta_1$ and $\beta_2$}
\tablehead{\colhead{} & 
\colhead{} & \colhead{} & \multicolumn{2}{c}{Isolated Halo} & \multicolumn{2}{c}{Composite Model} \\
\\ \colhead{Model} &
\colhead{$\beta_1 \left ({\rm Gyr}^{-1}\right )$} 
& {$\beta_2\left ({\rm Gyr^{-1}}\right )$} & {b} & {c} & {b} & {c} \\}
\startdata
1a & 0.12 & 0.04 & 0.93 & 0.86 & 0.95 & 0.85\\
1b & 0.08 & -0.04 & 0.86 & 0.93 & 0.88 & 0.90\\
1c & -0.08 & -0.12 & 0.93 & 1.07 & 0.93 & 1.06\\
2a & 0.16 & 0.16 & 1.00 & 0.78 & 1.01 & 0.77\\
2b & 0.0 & -0.16 & 0.78 & 1.00 & 0.80 & 0.99\\
3a & 0.24 & 0.08 & 0.89 & 0.73 & 0.92 & 0.73\\
3b & 0.16 & -0.08 & 0.73 & 0.89 & 0.75 & 0.87\\
3c & -0.16 & -0.24 & 0.82 & 1.12 & 0.82 & 1.08
\enddata
\end{deluxetable}

\end{document}